          [2001/04/25 1.1 (PWD)]
\documentclass[a4paper,twocolumn]{spaceDebrisC} % European paper
\usepackage{times}
\usepackage[numbers]{natbib}
\usepackage{graphicx}

\title{New SST Optical Sensor of Pampilhosa da Serra -- studies on image processing algorithms and multi-filter characterization of Space Debris.}
\author[1]{B. Coelho}
\author[1]{D. Barbosa}
\author[1,2]{M. Bergano}
\author[3]{A.C.M. Correia}
\author[4]{J. Freitas}
\author[5]{P. Marques}
\author[1,6]{J. Pandeirada}
\author[1]{V. Ribeiro}
\affil[1]{Instituto de Telecomunicações, 3810-193 Aveiro, Portugal, Email: brunodfcoelho$@$av.it.pt}
\affil[2]{ESTGA—Universidade de Aveiro, 3750-127 Aveiro, Portugal}
\affil[3]{CFisUC, Universidade de Coimbra, Coimbra, Portugal}
\affil[4]{MDN,1400-204 Lisboa, Portugal}
\affil[5]{Instituto de Telecomunicações/ISEL-IPL, 1049-001 Lisboa, Portugal}
\affil[6]{Departamento de Electrónica, Telecomunicações e Informática—Universidade de Aveiro, 3810-193 Aveiro,
Portugal}

\begin{document}


\maketitle

\begin{abstract}

As part of the Portuguese Space Surveillance \& Tracking (SST) System, two new Wide Field of View (2.3ºx2.3º) small aperture (30cm) telescopes will be deployed in 2021, at the Pampilhosa da Serra Space Observatory (PASO), located in the center of the continental Portuguese territory, in the heart of a certified Dark Sky area. These optical systems will provide added value capabilities to the Portuguese SST network, complementing the optical telescopes currently in commissioning in Madeira and Azores. These telescopes are optimized for GEO and MEO survey operations and besides the required SST operational capability, they will also provide an important development component to the Portuguese SST network. The telescopes will be equipped with filter wheels, being able to perform observations in several optical bands including white light, BVRI bands and narrow band filters such as H(alpha) and O[III] to study potential different object’s albedos. This configuration enables us to conduct a study on space debris classification$/$characterization using combinations of different colors aiming the production of improved color index schemes to be incorporated in the automatic pipelines for classification of space debris. This optical sensor will also be used to conduct studies on image processing algorithms, including source extraction and classification solutions through the application of machine learning techniques. Since SST dedicated telescopes produce a large quantity of data per observation night, fast, efficient and automatic image processing techniques are mandatory. A platform like this one, dedicated to the development of Space Surveillance studies, will add a critical capability to keep the Portuguese SST network updated, and as a consequence it may provide useful developments to the European SST network as well.
\end{abstract}

\section{Introduction}

The number of pieces of space debris — including broken pieces of artificial satellites and wreckage from rockets — has been increasing yearly with the progress of many country’s space sectors and
rise in satellite launches. This rising population of space debris increases the potential danger to all
space vehicles and in-space infrastructure, from expensive communications satellites, Earth
Observation satellite constellations, to the International Space Station, space shuttles and other
spacecraft with humans aboard. For this reason, it is important to create a set of preventive
measures in order to avoid any damages to space satellites.

A network of optical systems using telescopes with large fields of view (FOV) has proven to be a
most effective way of surveying and tracking the ever increasing debris population. In particular the
orbit occupancy in LEO and MEO orbit faces big challenges with planned deployment of space
mega-constellations requiring sensors with strong surveying and tracking capabilities. But also the
strategic GEO orbits do present a growing debris population. Since by default GEO objects are
stationary or slowly evolving in the sky, surveying sensors with large FOV and higher integration
times are better suited.

Clearly, information on the orbital debris environment is crucially needed to determine the current
and future hazards that orbital debris poses to space operations since the orbital environment is
dynamic and in constant change. Unfortunately, this environment is difficult to accurately
characterize since only the largest of debris can be repeatedly tracked by ground-based sensors.
Several studies have shown the advantage of multiple site data integration in improving the accuracy
of orbit prediction and the possibility of cataloging small debris. In this sense the complementary
between optical and radar sensors do offer a very high value service to deliver good performance in
monitoring and tracking debris of many sizes and in a wide orbit range.\citep{esa18} 

In this article we described an new optical sensor to be installed at PASO (Pampilhosa da Serra Space Observatory), which will contribute for the Portuguese SST network, may be also an valuable asset for the Portuguese participation in the EU-SST, and that will constitute a crucial platform for research and development(R\&D) of image processing algorithms and to conduct multi-filter studies for characterization of space debris.

\section{New SST Optical Sensor at PASO}

The New SST Optical Sensor to be installed at PASO site. In the center of Portugal, PASO is located in the municipality of Pampilhosa da Serra, where ATLAS the First Portuguese Radar Tracking Sensor for Space Debris (see Pandeirada et al. also in this conference proceedings). Being in the heart of a Dark Sky certified region, PASO location has excellent all sky clearance, in a hilltop at 840 m, and surrounded by mountain ranges that allow protection from the light pollution of the coastal regions of the country. The conditions are among the best in continental Portugal for Optical use, where sky background reaches mag 21 or above and with more than 200 clear nights per year.

\begin{figure}
\centering
%\vspace{4cm}
\includegraphics[width=0.95\linewidth]{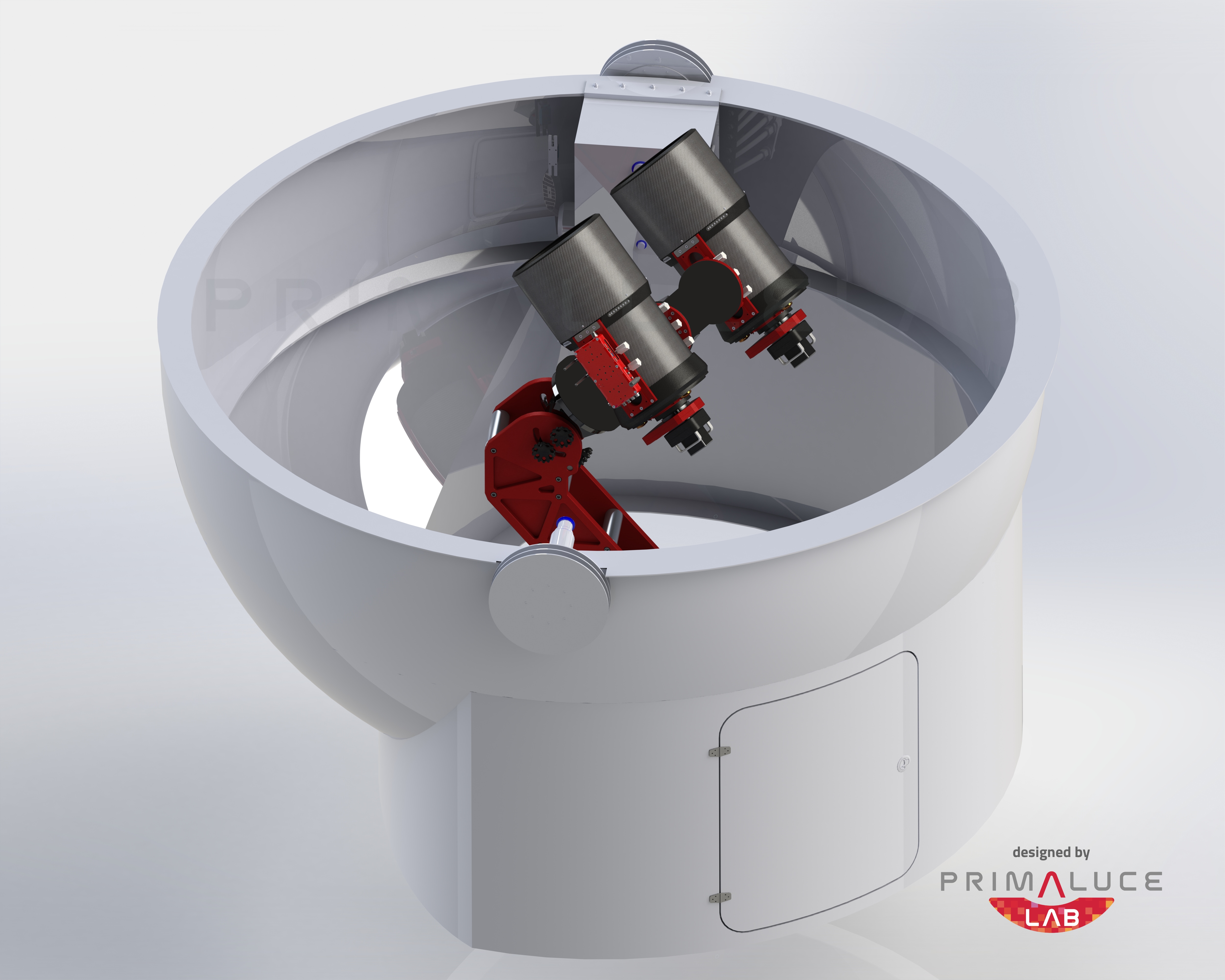}
\caption{Optical Sensor to be installed at PASO, consisting of two large FOV telescopes, equiped with filters B,V, R, I, H(alpha), O[III], here seen inside the dome. Image courtesy by PrimaLuceLab, the astronomical system provider and integrator. The EQ mount has a very fast speed enabling space debris surveillance.}\label{fig1single}
\end{figure}

This sensor will have dual use, SST (main focus) and Science. The sensor configuration with twin Optical Tube Assemblies (OTAs) with large FOV is optimized to operate in survey mode of Geosynchronous Equatorial Orbit (GEO) and Medium Earth Orbit objects (MEO) in the context of the Portuguese SST network. It will be able to perform observation in white light for SST operations, but being equipped with filters B, V, R, I, H(alpha) and O[III] the configuration capabilities are also fully suited for other uses, namely debris characterization studies, observing techniques development and scientific studies. It was also considered the flexibility of the system to accommodate future upgrades, if needed, and for example to develop the national capability towards LEOs observation in the future, through the use of a fast EQ mount. The system will be provided and integrated by PrimaLuceLab and can be seen figure \ref{fig1single}.

\begin{figure}
\centering
%\vspace{4cm}
\includegraphics[width=0.95\linewidth]{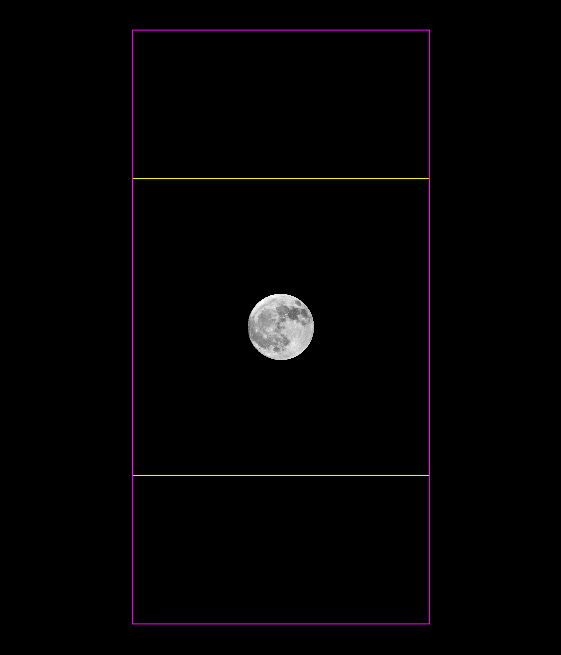}
\caption{FOV using the moon as a reference: in purple - a 4ºx2.3º field for the survey mode, in which the telescopes observe two different fields at same time; in yellow – a 2.3ºx2.3º FOV corresponding to each telescope, it can be also observed by the two telescopes at the same time to increase the signal to noise when performing observation with filters. (simulation obtained with astronomy.tools)}\label{fig2single}
\end{figure}

In figure \ref{fig2single} can be seen a simulation of the FOV of 4ºx2.3º obtained with the to telescopes observing contiguous fields at the same time, this configuration is optimized for maximum sky coverage is excellent for SST survey observation with white light. However, the system will allow also to use both telescopes to image the same field of 2.3ºx2.3º, this allows to increase the signal to noise when performing observations with filters, or even to do observation of a given source in two different filters at the same time. This twin OTA configuration also provides redundancy in the system, and constitutes an ideal test platform for the development of instrumentation such as cameras allowing comparison of devices in real similar conditions of observation.  

 \section{Studies on multi-filter characterization of space debris}
 A development component within the SST effort, is the possibility to conduct studies on space debris classification and characterization, using multi band measurements, and in this way trying to obtain color indexes that can be integrated in automatic classification processes, that can add extra, and useful information for SST operations. 
 
 There are some works that showed that multi band observations can be used to characterize Space Debris (eg. \citep{Cardona16}, \citep{Xiao16}, \citep{Lu17}). We intend to increase the samples of objects observed in order to investigate the possibility of statistically refine color indexes for automatic characterization of the objects, at the same time study light curves not only in white light but also in different filters, to infer rotation parterns.

\section{Studies on image processing algorithms}
 
The central product of an SST system is an object catalogue, which must contain updated orbit information for all detected objects by the entire network. The optical sensor proposed here will operate in Survey mode of GEOs and MEOs objects contributing for the Portuguese SST network, complementing the optical telescopes in Madeira and
Azores.

The image processing pipeline will constitute the first of the development fronts associated with the sensor. This is a hot topic in science, due to the modern and future large infrastructures that are being prepared at international level. The need for fast and effective image processing pipelines are taking techniques and computing solutions to new levels of performance and automation, adopting the principles of machine learning and AI.

\begin{figure*}
\centering
\vspace{4cm}
\includegraphics[width=1.0\textwidth]{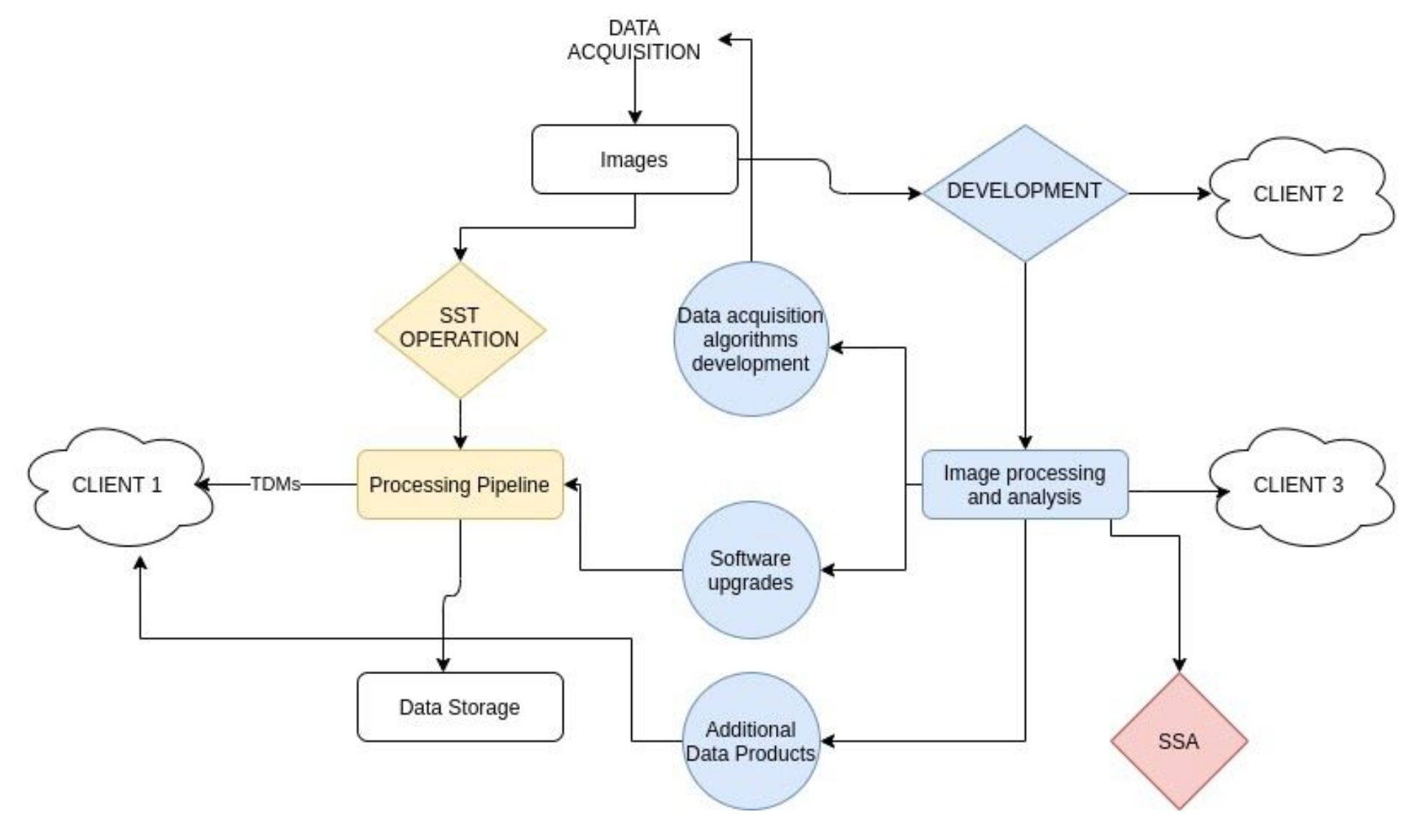}
\caption{Overview of the platform for development on image processing algorithms,
and to conduct studies on multi-filter characterization of Space Debris, here we
intend to increase the samples in terms of the number of objects, but also in terms of
time variability of the colors.\label{fig:double}}
\end{figure*}

In figure \ref{fig:double} is shown a scheme of the operation and development scenarios to be applied.

The data acquisition is made at the telescope as images that can be part of SST operations, or for development and research use. The results of the image processing obtained in SST operations are going to be sent to the Portuguese SST National Operation Center (client 1 in the figure 7). Images obtained for development and science can be processed and used for example for the studies of multi-filter classification of Space Debris. On the other hand we can obtain images for Science programs that can be sent directly to the client 2, or processed by observatory staff, with the results to be sent to the client 3. All these processes will involve interaction with the different clients, allowing pipelines and procedures to evolve to meet their needs, in the most effective way possible.

\section*{Acknowledgments}

%The team acknowledges financial support from ENGAGE-SKA Research Infrastructure, ref. POCI-01-0145-FEDER-022217, funded by COMPETE 2020 and FCT, Portugal; from the European Commission H2020 Programme under the grant agreement 2-3SST2018-20; exploratory project of reference IF/00498/2015 and PHOBOS project grant POCI-01-0145-FEDER-029932, funded by Programa Operacional Competitividade e Internacionalizacão (COMPETE 2020) and FCT, Portugal. IT team members acknowledge support from Projecto Lab. Associado UID/EEA/50008/2019.

The team acknowledges financial support from the European Commission H2020 Programme under the grant agreement 2-3SST2018-20.

\end{document}